\documentclass[
 amsmath, amssymb, breqn,
 aps, prb, superscriptaddress, twocolumn,
longbibliography]{revtex4-2}

\usepackage{graphicx}
\usepackage{amsmath,amssymb}
\usepackage{textcomp}
\usepackage{dcolumn}
\usepackage{hyperref}

\newcommand{\dd}{\ensuremath{\mathrm{d}}}

\newcommand{\pdiff}[2]{\ensuremath{\frac{\partial {#1}}{\partial {#2}}}}

\newcommand{\heeps}{\ensuremath{\varepsilon_\mathrm{He}}}
\newcommand{\modea}{{TE$_\mathrm{011}$}} 
\newcommand{\modeb}{{TE$_\mathrm{111}$}} 

\def\3He{$^3$He}
\def\4He{$^4$He}
\begin{document}

\title{Precision measurements of the zero temperature dielectric constant and density of liquid $^4$He}

\author{R. T. Learn}
\author{E. Varga}
\email{ev@ualberta.ca}
\author{V. Vadakkumbatt}
\author{J. P. Davis}
\email{jdavis@ualberta.ca}
\affiliation{Department of Physics, University of Alberta, Edmonton, Alberta, T6G 2E1, Canada}

\date{\today}

\begin{abstract}
The resonant frequencies of three-dimensional microwave cavities are explicitly dependent on the dielectric constant of the material filling the cavity, making them an ideal system for probing material properties. In particular, dielectric constant measurements allow one to extract the helium density through the Clausius-Mossotti relation. By filling a cylindrical aluminum cavity with superfluid helium, we make precision measurements of the dielectric constant of liquid $^4$He at saturated vapor pressure for range of temperatures 30 -- 300 mK and at pressures of 0-25.0 bar at 30 mK, essentially the zero temperature limit for the properties of $^4$He. After reviewing previous measurements, we find systematic discrepancy between low and high frequency determination of the dielectric constant in the zero-temperature limit and moderate discrepancy with previously reported values of pressure-dependent density. Our precision measurements suggest 3D microwave cavities are a promising choice for refining previously measured values in helium, with potential applications in metrology. 


\end{abstract}

\maketitle

\section{Introduction}

Three dimensional (3D) microwave cavities are an important tool for the physicist.  For example, they are used as accelerating cavities in particle colliders \cite{Padamsee1993} and are often used in combination with transmon qubits \cite{Reagor2016,Wang2016} --- one of the most promising qubit architectures.  One reason for the ubiquity of the 3D microwave cavity is that the open structure allows the electric and magnetic fields to reside in a material-free volume, reducing dissipation from lossy materials \cite{Reagor2013,Kudra2020}. This can be compared with on-chip microwave cavities, where substrate loss from two-level systems \cite{Bejanin2021,Muller2019,Martinis2005} generally dominates.  Furthermore, the open structure allows for the incorporation of materials into the microwave cavity \cite{Tretiakov2020,Ruether2021} making 3D cavities a valuable tool for precision measurements of material properties \cite{Hanson1976,Berthold1976a}.

Incorporating superfluid \4He with microwave systems is beneficial for multiple applications. Filling a 3D microwave cavity with superfluid helium allows easy tunability of the cavity frequency \cite{Souris2017}, and improves thermalization of superconducting qubits \cite{Lane2020}; superfluid helium, when coupled to a microwave optomechanical system, is a promising mechanical medium for proposed detectors of gravitational waves \cite{Lorenzo2014,Singh2017,Vadakkumbatt2021} and dark matter \cite{Manley2020}; and allows novel studies of 2D electron systems \cite{Zadorozhko2018}, including a design of new type of qubit \cite{Pollanen2021}.

Here, we use a 3D microwave cavity for a precision study of the dielectric constant and density of superfluid $^4$He in the low temperature limit. Using the Clausius-Mossotti relation, the dielectric constant measurements can be interpreted as measurements of the helium density, in a manner essentially similar to approaches such as dielectric constant (or refractive index) gas thermometry \cite{Rourke2021,Gaiser2018}. The ability to resolve small frequency shifts in the high-$Q$ microwave cavity allows this to be done with precision comparable to -- or exceeding -- the state-of-the-art capacitance measurements \cite{Kerr1964,Harris-Lowe1970,Niemela1995,Tanaka2000,Martinis2005}. Interestingly, for temperature dependence of the dielectric constant, we find a systematic discrepancy between low (capacitive) and high (microwave resonance) frequency determination of the dielectric constant which cannot be accounted for by frequency dependence of the polarizability. For a particular choice of polarizability of helium \cite{Gaiser2018}, we find good agreement with commonly used literature values \cite{Brooks1977,Abraham1970} of pressure dependence of the low-temperature density and the speed of sound. We find the largest sources of uncertainty to originate in the value of the molar polarizability of liquid helium and complex deformation of the cavity in a pressurized bath.

\section{Theory}
\begin{figure}
    \centering
    \includegraphics[width=1.0\columnwidth]{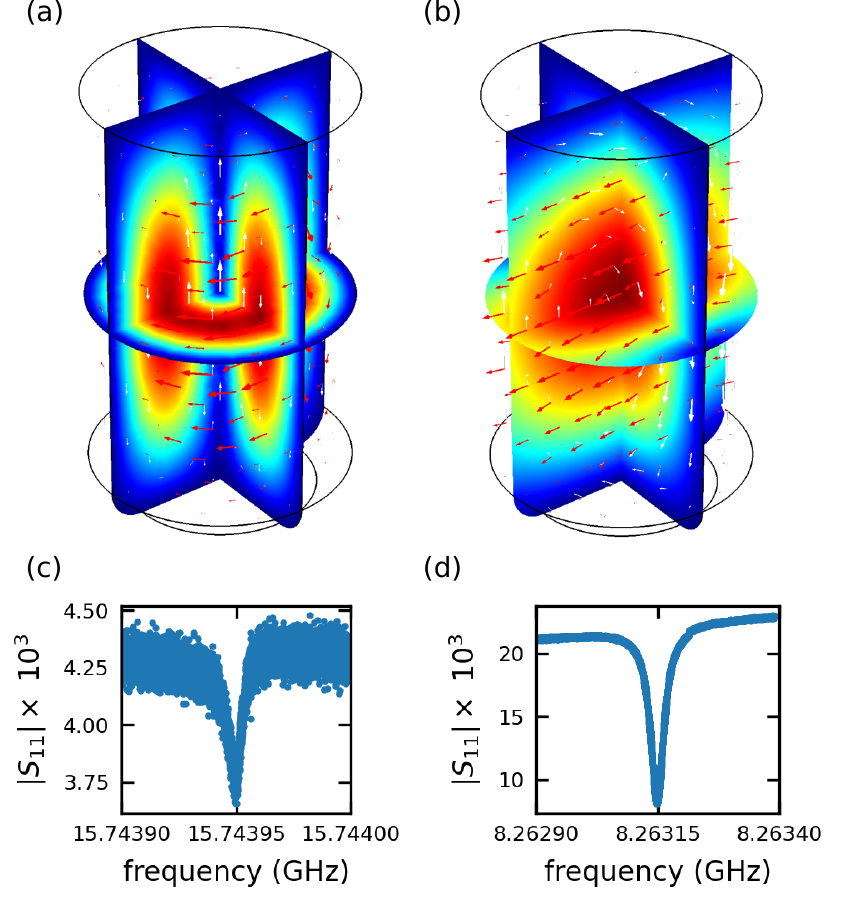}
    \caption{(a) Finite-element simulation  of  the  electric  field  for  the  TE$_{011}$  mode  of a right  cylindrical cavity with resonance frequency 15.693 GHz at room temperature. Red arrows show the direction of electric field, while white arrows show the direction of magnetic field. The lower edge of the cavity is filleted to split the desired TE$_{011}$ mode from the low-$Q$ degenerate TM$_{111}$ mode. (b) Simulation of the electric field for the TE$_{111}$ mode. (c) Example measurement for the TE$_{011}$ mode while in-vacuum at base temperature, with a total $Q$ of 2$\times 10^{6}$. (d) Example measurement for the TE$_{111}$ mode while in-vacuum at base temperature, with a total $Q$ of 1.6$\times 10^{5}$.}
    \label{fig:modes_fig}
\end{figure}

The helium density was calculated by measuring the resonant frequency of a cylindrical microwave cavity, which will have standing-wave modes determined by Maxwell's equations. For a right cylinder with height $h$ and radius $a$, the resonant frequencies for transverse electric (TE) modes, where $E_z=0$ with $z$ the axis of the cylinder, are given by \cite{Pozar2011}
\begin{equation}
    f_{\rm{nml}} = \frac{c}{2\pi\sqrt{\mu_{\rm{r}}\epsilon_{\rm{r}}}}\sqrt{\left(\frac{x'_{\rm{nm}}}{a}\right)^{2}+\left(\frac{l\pi}{h}\right)^{2}},
\label{eq:TE_modes}
\end{equation}
where $x'_{nm}$ is the $m$th zero of the derivative of the $n$th Bessel function of the first kind, $c$ is the speed of light, and $\mu_{\rm{r}}$, $\epsilon_{\rm{r}}$ are the relative permeability and permittivity, respectively, of the material filling the cavity. In particular, we are interested in the \modea{} and \modeb{} modes, pictured in Fig.~\ref{fig:modes_fig}. Particularly for the mode \modea{}, since the electric field vanishes at all surfaces for this mode, dielectric and seam losses are negligible, and only conductor losses contribute \cite{Reagor2015}. This results in a low loss rate and hence a high-quality mode, capable of achieving internal quality factors on the order of 10$^{8}$ for high-purity aluminum cavities \cite{Reagor2013}.

When filling a microwave cavity with superfluid helium, only the relative permittivity of the material inside the cavity changes, and the filled resonant frequency $f_{\rm{filled}}$ will be reduced from the in-vacuum resonant frequency $f_{\rm{empty}}$ by the relation 
\begin{equation}
    f_{\rm{filled}}=\frac{f_{\rm{empty}}}{\sqrt{\epsilon_{\rm{He}}}},
    \label{eq:filled-empty}
\end{equation}
where $\epsilon_{\rm{He}}$ is the dielectric constant of superfluid \4He. This allows us to directly calculate the dielectric constant by comparing measurements of the filled and in-vacuum resonant frequencies as
\begin{equation}
    \label{eq:eps-pressure-uncorrected}
    \heeps = \left(\frac{f_\mathrm{empty}}{f_\mathrm{filled}(P)}\right)^2.
\end{equation}
This expression, however, needs to be corrected for finite compressiblity of the cavity walls (see \eqref{eq:eps-pressure-corrected} below).

The dielectric constant and density are related through the Clausius-Mossotti relation
\begin{equation}
    \frac{\heeps-1}{\heeps+2}=\frac{4\pi}{3}\frac{\alpha}{M}\rho,
\label{eq:clausius-mossotti}
\end{equation}
where $\alpha$ is the polarizability volume per mole (shortened to `polarizability' henceforth), $M$ is the molar mass of \4He and $\rho$ is the density \footnote{The polarizability volume per mole $\alpha$ is related to the molecular polarizability (per mole) $p = 4\pi\varepsilon_0 \alpha$. In dielectric constant gas thermometry literature (e.g., \cite{Gaiser2018}) the common definition of polarizability is $A_\varepsilon = 4\pi\alpha/3$.}.

Finally, given a pressure dependence of the density, the long-wavelength limit of the speed of sound is
\begin{equation}
    \label{eq:sound-speed}
    c = \left(\pdiff{\rho}{p}\right)^{-1/2}.
\end{equation}

\section{Experimental setup}
\label{experiment}

\begin{figure}
    \centering
    \includegraphics{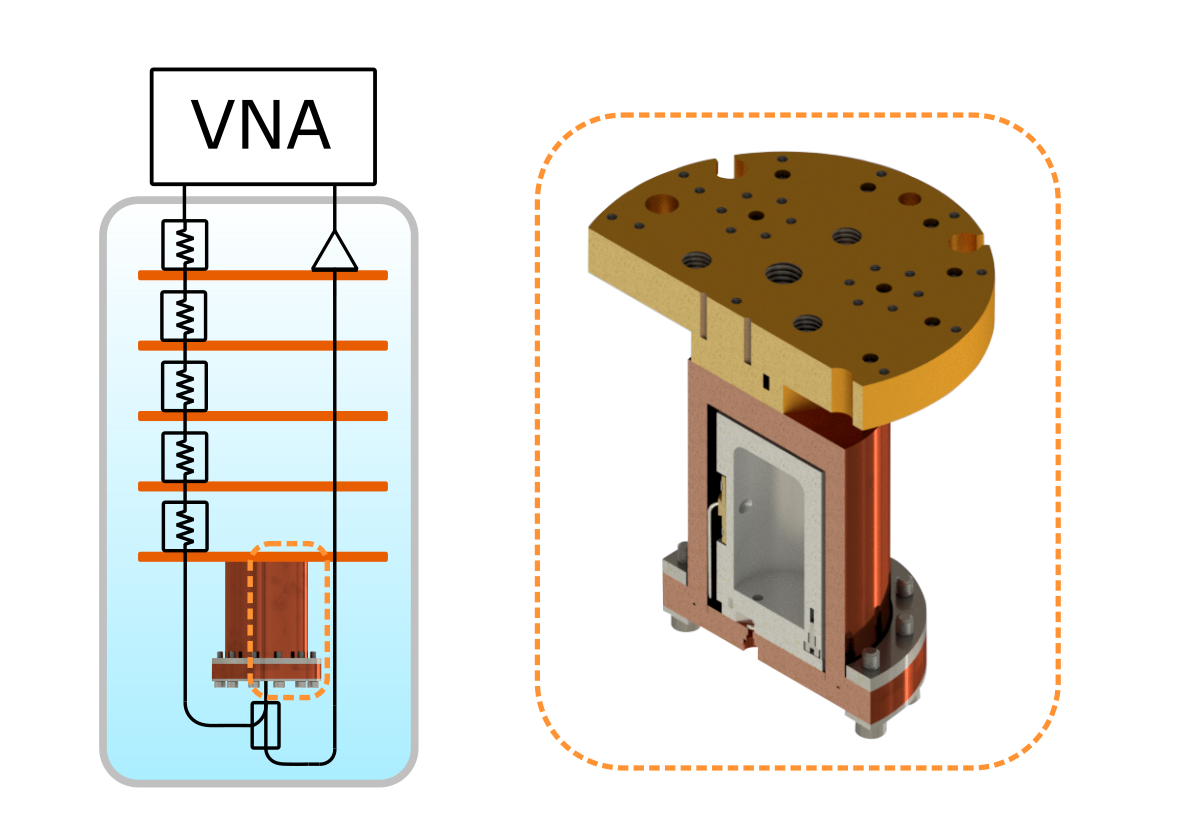}
    \caption{(a) Diagram of the circuit used to take measurements. A total of 34 dB of attenuation reduces the signal from the VNA before reaching the cavity, including the directional coupler which transmits and receives the $S_{11}$ signal. A low noise HEMT amplifies the signal at the 4K stage before returning to the VNA. (b) Cross-section of the experimental apparatus used to measure the dielectric constant of helium-4. The microwave cavity is placed inside a hermetically sealed cell, which is then filled with helium, preventing pressure differentials across cavity walls. }
    \label{fig:schematic}
\end{figure}

The dielectric constant was measured using a superconducting cylindrical aluminum microwave cavity cooled to approximately 30 mK using a dilution refrigerator, as schematically shown in Fig.~\ref{fig:schematic}. The cavity was machined to be 2.4 cm in diameter and 4 cm tall at room temperature, for an approximate total volume of 18 cm$^{3}$. The cavity was designed to operate in the \modea{} and \modeb{} modes, with a measured resonant frequencies of 15.74 GHz and 8.26 GHz, respectively, for $T=30$ mK. For our cavity machined out of 6061 aluminum, we measured an internal quality factor at 30 mK of $\sim$ 2 $\times$ 10$^{6}$ for \modea{}. Higher quality factors could be achieved with pure aluminum or niobium \cite{Reagor2013,Allen1971}. The \modea{} mode of a perfectly cylindrical cavity is degenerate with the low-$Q$ TM$_{111}$ mode. To break this degeneracy, a 1-inch-radius fillet was added to the bottom edge of the cavity. The microwave cavity was placed inside a hermetically sealed copper cell, as shown in Fig.~\ref{fig:schematic} (b). This design allows helium to freely flow in and around the microwave cavity such that there are no pressure differentials across the walls of the cavity. This eliminates the possibility of the cavity bowing under high pressures, which can shift the resonant frequency of the cavity considerably, as was seen in past measurements \cite{Souris2017}.

The complex-valued reflection from the cavity is fit to the expected frequency dependence of the scattering parameter \cite{Probst2015}
\begin{equation}
    \label{eq:S11}
    S_{11}(f) = Ae^{i(f\tau + \phi_0)}\left(1 - \frac{2e^{i\phi_Z}\frac{Q_\mathrm{tot}}{Q_\mathrm{ext}}}{1 + 2iQ_\mathrm{tot}\left(\frac{f}{f_0} - 1\right)}\right),
\end{equation}
where $f_0$ is the resonance frequency, $\phi_Z$ characterizes impedance mismatch and $A$, $\tau$ and $\phi_0$ characterize the overall loss, delay and phase rotation due to wiring and amplification within the cryostat \cite{Probst2015}. The total quality factor is given by $Q_\mathrm{tot}^{-1} = Q_\mathrm{ext}^{-1} + Q_\mathrm{int}^{-1}$ where $Q_\mathrm{ext}$ is the external quality factor characterizing the coupling to the microwave mode and $Q_\mathrm{int}$ is the internal quality factor due to all other dissipation processes \cite{Probst2015}.

The cavity was coupled to using a pin coupler, which was aligned parallel to the electric field of the TE$_{011}$ mode. By adjusting the length of the pin coupler, we decreased the coupling such that the external quality factor was $\sim 3 \times 10^{7}$ at base temperature. This meant that the total cavity quality factor was almost entirely limited by internal losses, allowing for maximum precision in our measurements.

To take measurements in the zero-temperature limit, the cell was mounted on the mixing chamber plate of a dilution refrigerator. The resonant modes were measured using a vector network analyzer (VNA), in an RF circuit shown in Fig.~\ref{fig:schematic} (a). At each stage of the dilution refrigerator, attenuators were used to heat sink the microwave coaxial line, attenuating the signal a total of 24 dBm. A directional coupler with 10 dB of attenuation was used to transmit the microwave power to and from the cavity. The reflected signal was then amplified through a low noise amplifier at the 4K stage of the refrigerator and returned to the VNA. 

The temperature of the mixing chamber was measured using an ultra-low-temperature ruthenium oxide sensor and controlled through an AC resistance bridge. Close to the base temperature, we were able to achieve temperature stability of 0.5 mK, while at higher temperatures ($T \approx 300$ mK), temperature stability decreased to 1 mK. The pressure of the helium inside the cell was set using a homemade control system, which consists of a ballast volume immersed in liquid nitrogen with a resistive heater controlled by a PID loop. The stability of the pressure measured at room temperature was about 1 mbar.

Once at base temperature ($T\approx30$ mK), measurements of both \modea{} and \modeb{} modes were taken in vacuum over several days to ensure that the resonant frequency was stable. The resonant frequency was observed to shift no more than a few hundred Hz (i.e., less than 1 ppm), which is within the precision of the fitting method used. The cavity was then filled with helium, and measurements were taken first at saturated vapor pressure (SVP), and then pressurized up to 25 bar. For each pressure, 200 traces of the $S_{11}$ signal were taken for both modes, which were individually fit using \eqref{eq:S11} to find the resonant frequency and its standard deviation. The pressurized density was measured during two cool downs of the dilution refrigerator. These data sets will be referred to as Run 1 and Run 2.

\section{Drive Power Analysis}
To ensure that the drive power would not heat the cavity, the effect of the VNA drive power was measured for the cavity while in-vacuum, filled to SVP, and pressurized to 25.0 bar. In each set of measurements, the drive power at source was varied from -20 to 0 dBm (0.01 to 1.00 mW), and the resonant frequency of the \modea{} mode was measured.

\begin{figure}
    \centering
    \includegraphics{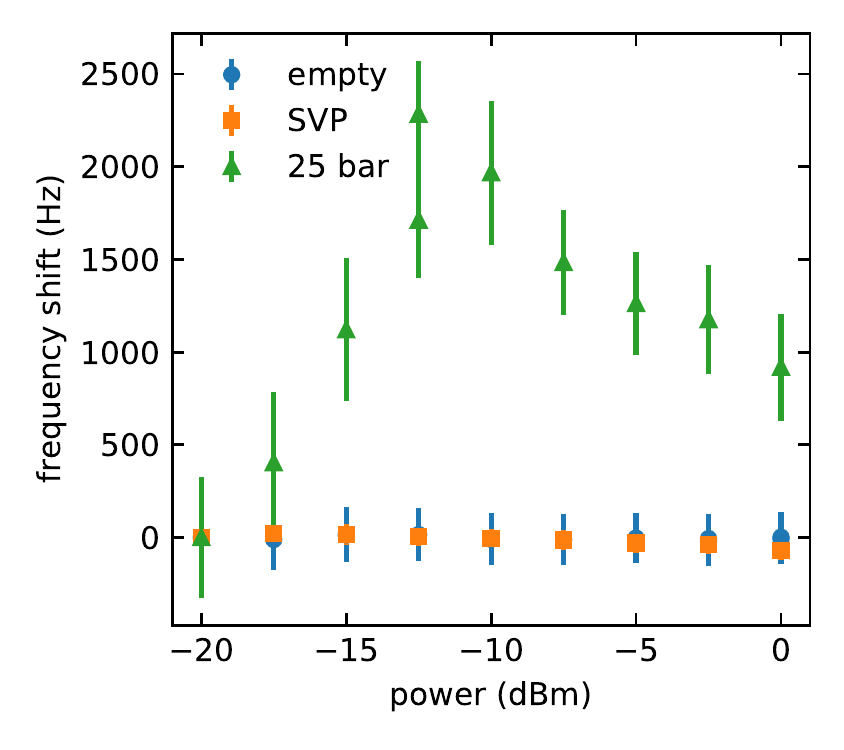}
    \caption{The shift in frequency of the \modea{} mode as the power (at source) is decreased from 0 to -20 dBm, for three separate sets of measurements: in-vacuum (blue), filled with helium at saturated vapor pressure (orange), and pressurized to 25 bar (green). The zero-point is taken as the -20 dBm for each respective dataset.}
    \label{fig:power_fig}
\end{figure}

Fig.~\ref{fig:power_fig} shows the resonant frequency of the \modea{} mode for powers between 0 and -20 dBm while in-vacuum, filled to SVP, and pressurized to 25 bar. Heating of the cavity ought to correspond to decrease of the resonance frequency \cite{Reagor2015}, which is observed weakly for the empty and SVP data. The origin of the peak in frequency shift for the 25 bar data is unknown, but it is unlikely to be related to the heating of the cavity itself. Nevertheless, even at the peak, the relative frequency shift is small and we use -10 dBm drive power which we find to be a good compromise between low-pressure heating and signal-to-noise ratio. However, to account for power-dependent frequency shifts we include an additional 100 Hz error on resonance frequencies measured at saturated vapour pressure (the data in Sec.~\ref{sec:res:eps-Tdep}) and additional 2500 Hz error on all resonance frequencies measured at increased pressures (the data in Sec.~\ref{sec:res:density}).

\section{Corrections due to systematic errors}
\label{sec:corrections}

The low temperature density of \4He was measured in the range 0.5-25.0 bar. While at 30 mK, the cell was pressurized in steps of 1 bar. Measurements were taken while ramping up the pressure from 1.0-25.0 bar, and then ramping down the pressure from 24.5-0.5 bar. There are multiple known sources of systematic error related to the pressure that may affect our measurements. Here we identify and correct for the following: 1) a hydrostatic pressure head, 2) superfluid fountain pressure, and 3) compression of the aluminum cavity.

\begin{enumerate}
    \item 
A hydrostatic pressure head arises from excess liquid helium in the fill line a height $h$ above the microwave cavity. This will not affect the measured frequency, but will shift the pressure in the cell $P_{\rm{cell}}$ from what is measured at room temperature $P_{\rm{meas}}$ to
\begin{equation}
    P_{\rm{cell}} = P_{\rm{meas}} + \rho g h
    \label{eq:hydrostat}
\end{equation}
where $g$ is the acceleration due to gravity.

Since the level of helium in the dewar surrounding the dilution unit is not constant, the level of liquid helium in the cell fill line will fluctuate, changing the hydrostatic pressure head. We estimate that the liquid level will vary between 59 and 79 cm above the cavity. We correct our data for a hydrostatic pressure head of height $h = 68.5$ cm, but consider the 59-79 cm range as one component of the uncertainty of the pressure reading.

\item 
When two reservoirs of He-II at different temperatures are connected via a thin channel that does not admit the flow of the viscous normal fluid component, a pressure difference develops according to
\begin{equation}
    \label{eq:fountain}
    \nabla p = \rho S\nabla T,
\end{equation}
where $S$ is the specific entropy \cite{Tilley_book}. The helium fill line of the cell passes through sintered copper heat exchangers on each stage of the dilution refrigerator that strongly restrict the flow of helium. Assuming that the effect is negligible below the temperature of the still ($T_\mathrm{still}\approx$~0.8~K) we assume that the pressure drop is dominated by the temperature gradient across the heat exchanger at the 1~K pot ($T_\mathrm{pot}\approx$ 1.45~K), i.e.,
\begin{equation}
    \label{eq:dp-fountain}
    \Delta p_\mathrm{fountain} = \int_{T_\mathrm{still}}^{T_\mathrm{pot}}\rho(p,T) S(p,T) \dd T,
\end{equation}
which has to be subtracted from the pressure measured at room temperature to obtain the correct cell pressure. Here, for $\rho(p,T)$ and $S(p,T)$, HEPAK dataset was used \cite{HEPAK} and the influence of the pressure gradient on the material parameters $\rho$ and $S$ was neglected. The resulting fountain pressure correction is shown in Fig.~\ref{fig:fountain-pressure}.

\begin{figure}
    \centering
    \includegraphics{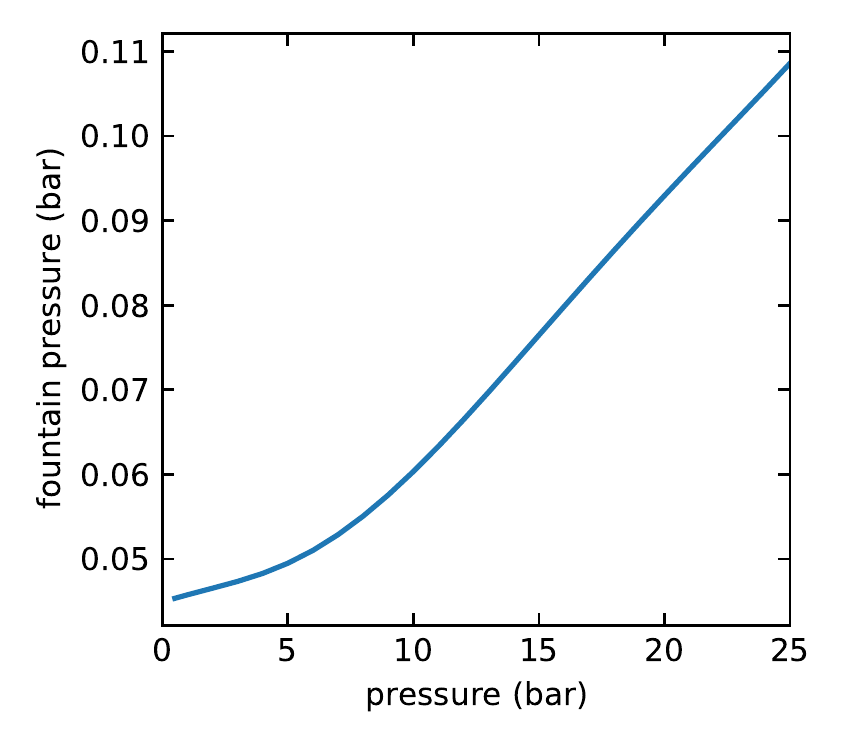}
    \caption{Pressure difference between two ends of a superleak due to the fountain effect calculated using \eqref{eq:dp-fountain}. End temperatures are assumed to be 0.8 K and 1.45 K.}
    \label{fig:fountain-pressure}
\end{figure}

\item
 As the pressure of the helium increases, the cavity frequency will also be affected by deformation of the cavity itself. Assuming small elastic deformation, all cavity dimensions $d$ (i.e., radius $a$ and height $h$) will be re-normalized to $d = d_0(1 - P/3K)$ where $P$ is the pressure, $K$ the bulk modulus of the cavity material and $d_0$ is the cavity dimension at $P=0$. The dielectric constant corrected for deformation is thus (cf.~Eq.~\eqref{eq:TE_modes})
\begin{equation}
    \label{eq:eps-pressure-corrected}
    \heeps = \left(\frac{f_\mathrm{empty}}{f_\mathrm{filled}(P)}\right)^2\frac{1}{(1 - P/3K)^2}.
\end{equation}
The bulk modulus is related to the Young's modulus $E$ through $K = E/3(1-2\nu)$, where for Al the Poisson ratio $\nu= 0.33\pm 0.01$. The value of the Young's modulus $E$ of Al alloy 6061 at low temperatures reported in the literature varies in the range \cite{Ekin2006,VanSciver2012} 77.7 -- 78 GPa. Error estimates are generally not available. Conservatively, we chose $E=77.8 \pm 0.5$ GPa. The uncertainties of pressure, $E$, and $\nu$ are propagated into the uncertainties $\varepsilon$ shown below. Yield stress of 6061 is in the range of 350 MPa \cite{Ekin2006}, making corrections due to plastic deformation under our highest applied pressure, approximately 2.5 MPa, negligible.

\end{enumerate}

Finally, a significant source of uncertainty for calculating the density from the measured frequency shift and dielectric constant using \eqref{eq:clausius-mossotti} is the value of the polarizability $\alpha$ of liquid \4He. Various values for the molar polarizability have been obtained \cite{Kerr1970,Kierstead1976a,Harris-Lowe1970,HEPAK,Berthold1976a} which disagree on the level of about 0.1\%. For helium in the gas phase, the value of the polarizability is known to satisfactory precision with ppm-level agreement between experiment \cite{Gaiser2018} and \emph{ab-initio} theory \cite{Piszczatowski2015}. However, due to inter-atomic interactions, the Clausius-Mossotti equation requires a virial expansion \cite{Buckingham1955,Hill1958}, suggesting that the value of effective polarizability of the liquid used in \eqref{eq:clausius-mossotti} will likely differ from the polarizability of individual atoms \cite{Kim2005,Gaiser2008a}.

Despite these difficulties, we believe the currently available low-density value of polarizability is the most reliable. In the following, we calculate density and speed of sound using the polarizability measured with a dielectric constant gas thermometer near the triple point of water \cite{Gaiser2018}, which produces the best absolute agreement with previously accepted values of density \cite{Abraham1970,Brooks1977} without the necessity of explicitly correcting using external reference \cite{Tanaka2000}.

\section{Results}

\subsection{Temperature dependence of the dielectric constant at saturated vapor pressure}
\label{sec:res:eps-Tdep}

The temperature dependence of the dielectric constant has been measured in the past, but few have measured it in the low temperature limit. Chan {\it et al.}~\cite{Chan1977} measured the dielectric constant using a parallel plate capacitor in the range 100-1200 mK. They obtained their value for the dielectric constant at zero temperature $\epsilon_{0}$ by extrapolating their data using
\begin{equation}
    \epsilon_{\rm{He}} = \epsilon_{0} + (\epsilon_{\rm{He}} - 1)(A_{4}T^{4} + A_{6}T^{6}).
    \label{eq:Chan_fit}
\end{equation}
Only Berthold {\it et al.}~\cite{Berthold1976a} have directly measured the dielectric constant at SVP in the low temperature limit. Their method was similar to ours --- they measured the resonant frequency of the TE$_{011}$ mode of a cylindrical niobium cavity. They do not report an exact temperature, saying only that their measurement was taken below 100 mK. The most recent measurement of the dielectric constant was made by Niemela \& Donnelly~\cite{Niemela1995}, who only measured the dielectric constant above 1 K, but extrapolated to zero temperature using empirical formulas.

We measured the dielectric constant at saturated vapor pressure (SVP) during Run 2 using the \modea{} mode. This was done by slowly filling the cell while watching the frequency shift using the VNA. Once the frequency stabilized, the cell was assumed to be filled, and the filling was halted by closing a room-temperature valve in the gas handling system. This method leads to some uncertainty in the height of the helium above the cell, which would increase the pressure in the cell from SVP by an unknown amount. Assuming the pressure head is no more than 1 cm above the cavity, the pressure would be increased by at most $10^{-4}$ bar. The temperature was increased in steps of 25 mK from 30-330 mK, and the shift in the resonant frequency was measured. This data was then compared to in-vacuum data at the same temperature through Eq.~\ref{eq:filled-empty}.

\begin{figure}
    \centering
    \includegraphics{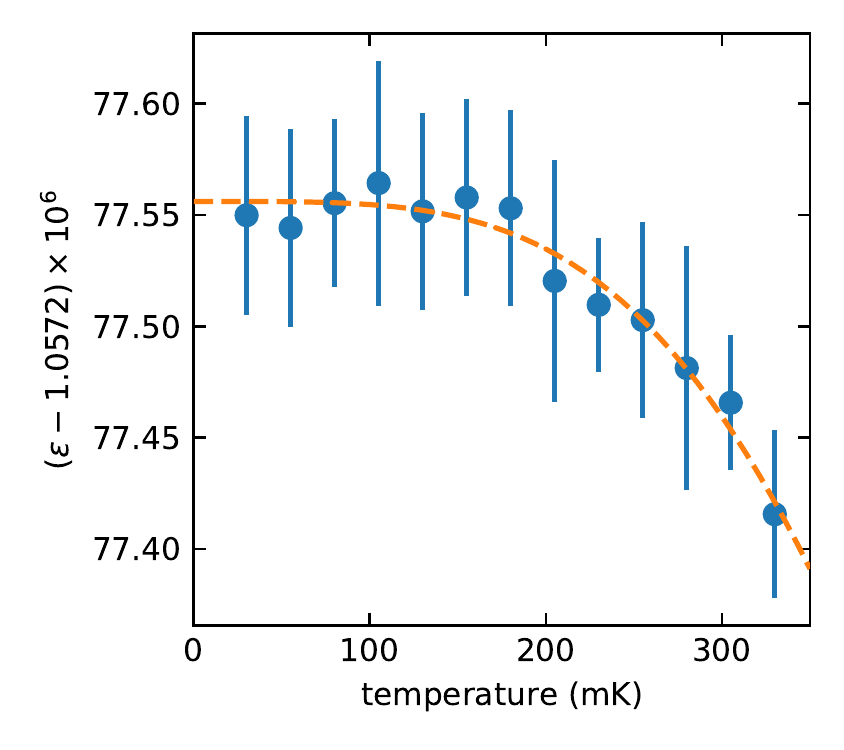}
    \caption{The dielectric constant of helium at saturated vapour pressure for temperatures between 30-330 mK measured using the \modea{} mode and calculated using \eqref{eq:filled-empty}. The error bars were calculated using the statistical errors on $f_\mathrm{filled}$ and $f_\mathrm{empty}$, which were obtained by measuring the cavity resonance 200 times. The dashed line is a fit to temperature dependence given by \eqref{eq:Chan_fit}. For the dielectric constant obtained using the \modeb{} mode (not shown), the temperature dependence was obscured by increased scatter but is consistent with \modea{} within approximately 3 ppm.}
    \label{fig:SVP_fig}
\end{figure}

Figure~\ref{fig:SVP_fig} plots our calculated dielectric constant at SVP as a function of temperature. Following Chan \cite{Chan1977}, we fit our data to Eq.~\ref{eq:Chan_fit} (orange dashed line in Fig.~\ref{fig:SVP_fig}), finding the zero-temperature dielectric constant $\varepsilon_0$ and the fitting parameters $A_4$ and $A_6$ to be
\begin{align*}
    \epsilon_0 &= 1.05727755 \pm 2\times 10^{-8}\\
    A_4 &= (-3 \pm 2)\times 10^{-4}\; \mathrm{K}^{-4}\\
    A_6 &= (1 \pm 2)\times 10^{-3}\; \mathrm{K}^{-6}.
\end{align*}
The uncertainties were estimated using a Monte-Carlo method, where multiple datasets were generated by drawing a random sample for each temperature from a normal distribution centered on the experimental mean and with variance equal to the square of the experimental error estimate. Each generated dataset was fit using \eqref{eq:Chan_fit}; the values shown above are the averages and standard deviations of a set of the individual fit parameters. The procedure was repeated for sufficiently high number of samples such that the estimates of values and their error have converged. The large uncertainties on $A_4$ and $A_6$ are due to weak temperature dependence of $\varepsilon$ in the range accessible to the present experiment, which results in a poorly conditioned fit.

Table~\ref{table:epsilon_0} summarizes the values for the zero-temperature dielectric constant obtained by several studies. Kierstead~\cite{Kierstead1976a} and Harris-Lowe and Smee~\cite{Harris-Lowe1970} extrapolated high temperature data to obtain a zero-temperature value. Note in Tab.~\ref{table:epsilon_0} that the value of zero-temperature dielectric constant obtained here is systematically higher, and outside of estimated error bars, than previously reported values measured using helium-filled capacitors but in excelled agreement with ref.~\cite{Berthold1976a}, the only other work measured using a microwave cavity. The reason for this relatively large and apparently systematic discrepancy between low-frequency and high-frequency estimation of the dielectric constant is at present unknown, since the relative change in polarizability between DC and 15~GHz is expected to be negligible, on the order of $10^{-12}$ \cite{Piszczatowski2015}.

\begin{table} 
    \begin{ruledtabular}
    \begin{tabular}{lcl}
      Authors & method & $\epsilon_{0}$\\
      \hline
      Niemela and Donnelly \cite{Niemela1995} & capacitance  & 1.057255 \\
      Chan {\it et al.} \cite{Chan1977} & capacitance & 1.0572190(5) \\
      Kierstead~\cite{Kierstead1976a}\footnote{Extrapolated to zero-temperature by Chan {\it et al.}} & capacitance & 1.0571374(10) \\
      Harris-Lowe and Smee~\cite{Harris-Lowe1970}$^{\rm{a}}$ & capacitance & 1.0572467(100) \\
      Tanaka {\it et al.}~\cite{Tanaka2000} & capacitance & 1.0572025 \\
      Berthold {\it et al.} \cite{Berthold1976a} & cavity & 1.0572784(5) \\
      Current authors & cavity & 1.05727755(2)\\
    \end{tabular}
    \end{ruledtabular}
    \caption{Various values for the dielectric constant of \4He at saturated vapor pressure extrapolated to $T=0$. The column 'method' shows whether the experiment measured a capacitance change of a helium-filled capacitor or or a shift of resonant frequency of a microwave cavity (as in the present experiment).}
    \label{table:epsilon_0}
\end{table}

\subsection{Pressure dependence of dielectric constant and density.}
\label{sec:res:density}

\begin{figure}
    \centering
    \includegraphics{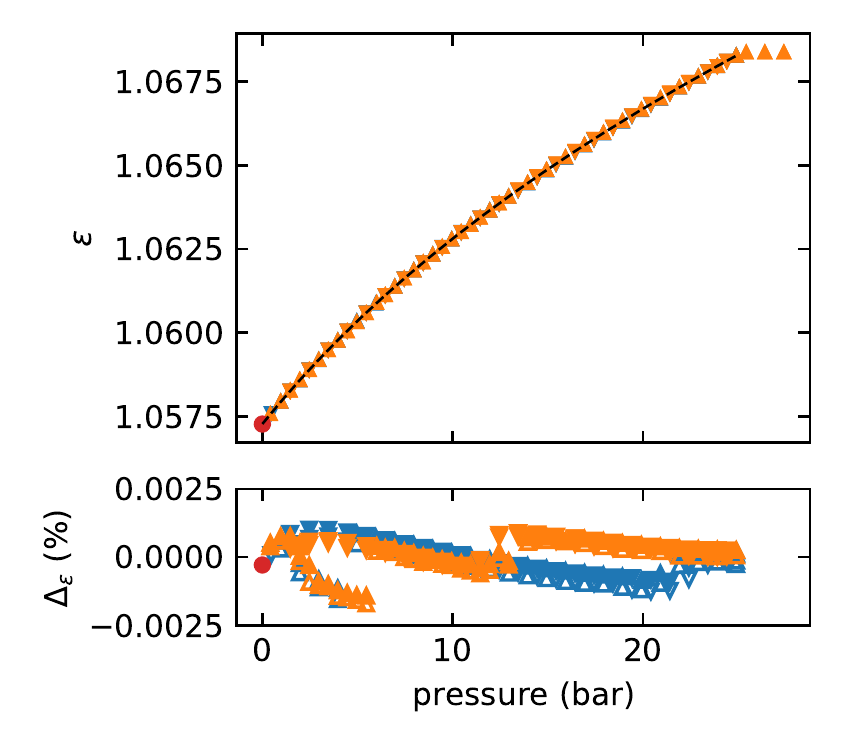}
    \caption{Pressure dependence of the dielectric constant (top) and the residuals from the fit \eqref{eq:eps-fit} (bottom). The data above 25 bar, where helium solidifies, were excluded from the fit. Up-triangles correspond to increasing pressure ramp, down-triangles to decreasing pressure; orange points to run 1 and blue points to run 2; full points to mode \modea{} and empty points to \modeb{}. The red circle shows the dielectric constant at the saturated vapour pressure.}
    \label{fig:eps}
\end{figure}

The corrected results for the low-temperature pressure-dependent dielectric constant are shown in Fig.~\ref{fig:eps}. This includes data taken over two separate runs, using two different microwave modes, increasing and decreasing pressure ramp, and a separate measurement at saturated vapour pressure.

Since the calculation of density from the dielectric constant depends on polarizability, which is not presently known with sufficient accuracy for liquid helium, we adopt an approach common in dielectric constant gas thermometry \cite{Gaiser2018} and fit the pressure dependence of the Clausius-Mossotti parameter $\mu = (\varepsilon - 1)/(\varepsilon + 2) \propto \rho$ (the proportionality assumes that the polarizability is density-independent) using a 3rd degree polynomial
\begin{equation}
    \label{eq:eps-fit}
    P = A_0 + A_1\mu + A_2\mu^2 + A_3\mu^3,
\end{equation}
where $A_n$ are fit parameters. To account for measurement errors in both pressure and dielectric constant, \eqref{eq:eps-fit} was fit to the data using orthogonal distance regression weighted by the estimated uncertainties of the individual measurements of pressure and $\mu$ \cite{Brown1990}. The errors of the fit parameters were estimated by bootstrap \cite{Efron1981}. The resulting parameter values are
\begin{align*}
    A_0 &= -114 \pm13~\mathrm{bar}\\
    A_1 &= (2.11\pm0.19)\times 10^4~\mathrm{bar}\\
    A_2 &= (-1.518\pm0.095)\times 10^6~\mathrm{bar}\\
    A_3 &= (3.81\pm0.15)\times 10^7~\mathrm{bar}.
\end{align*}
To invert \eqref{eq:eps-fit} and obtain the $\mu(P)$ (and $\varepsilon(P)$) relationship a standard root-finding algorithm is employed. The relative residuals $\Delta_\varepsilon = (\varepsilon - \varepsilon_\mathrm{fit})/\varepsilon_\mathrm{fit}$ are shown in the bottom panel of Fig.~\ref{fig:eps}. In the plot of the residuals, there are two distinct jumps in the data. The first jump occurs in both runs at $P \sim$ 5 bar while increasing the pressure, but does not follow the same behavior while decreasing the pressure. The second jump happens at higher pressures, $P\sim$ 13 and 21 bar for Run 1 and Run 2, respectively, and the jump in resonant frequency is observed both while increasing and decreasing the pressure. By taking multiple data sets, we have shown that these jumps are repeatable at similar pressures.

We suspect that these jumps are caused by mechanical slipping - the cavity lid or pin coupler may shift at certain pressures, causing the resonant frequency to shift since the effective volume of the cavity has changed. Berthold {\it et al.}~\cite{Berthold1976a} observe a similar effect, stating that mechanical shock arising from the opening and closing of valves in their system can cause a frequency shift of up to 1 kHz in their cylindrical microwave cavity.

\begin{figure}
    \centering
    \includegraphics{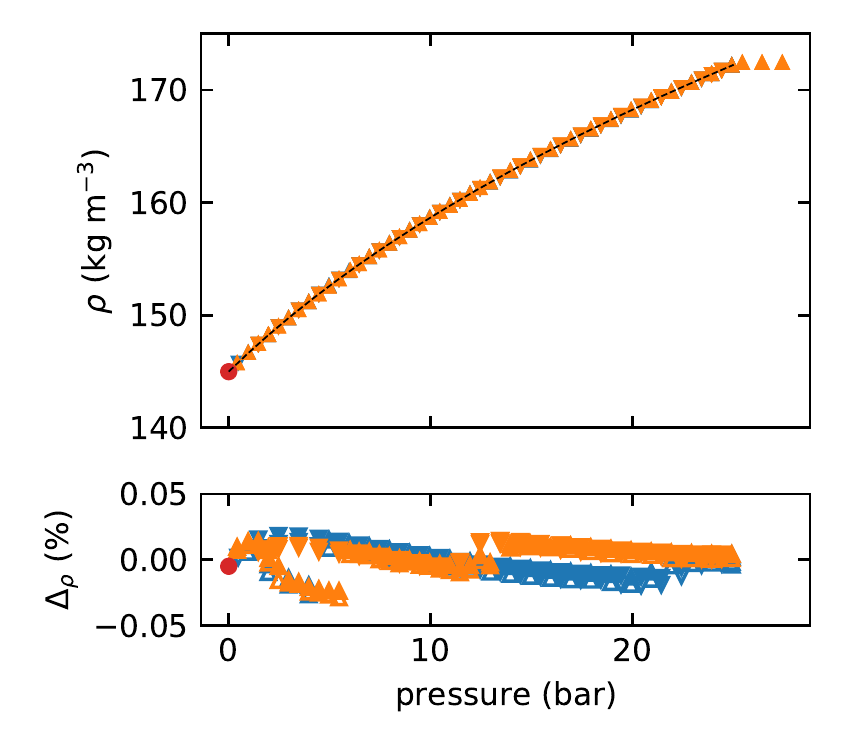}
    \caption{\4He density at 30 mK calculated using the Claussius-Mossotti relation \eqref{eq:clausius-mossotti} with polarizability $\alpha = 0.1234853$ cm$^{3}$ of Ref.~\cite{Gaiser2018} and the deviation from the dielectric constant fit \eqref{eq:eps-fit}. Styles of points as in Fig.~\ref{fig:eps}.}
    \label{fig:density}
\end{figure}

We calculate the density from $\mu$ with Clausius-Mossotti equations \eqref{eq:clausius-mossotti} using the polarizability determined through a helium-based dielectric constant gas thermometer near the triple point of water $\alpha = 0.1234853$~cm$^{3}$ \cite{Gaiser2018}. The resulting density and relative residuals are shown in Fig.~\ref{fig:density}. The residuals are calculated with respect to the density calculated from the fit of $\varepsilon$ \eqref{eq:eps-fit}.

\begin{figure}
    \centering
    \includegraphics{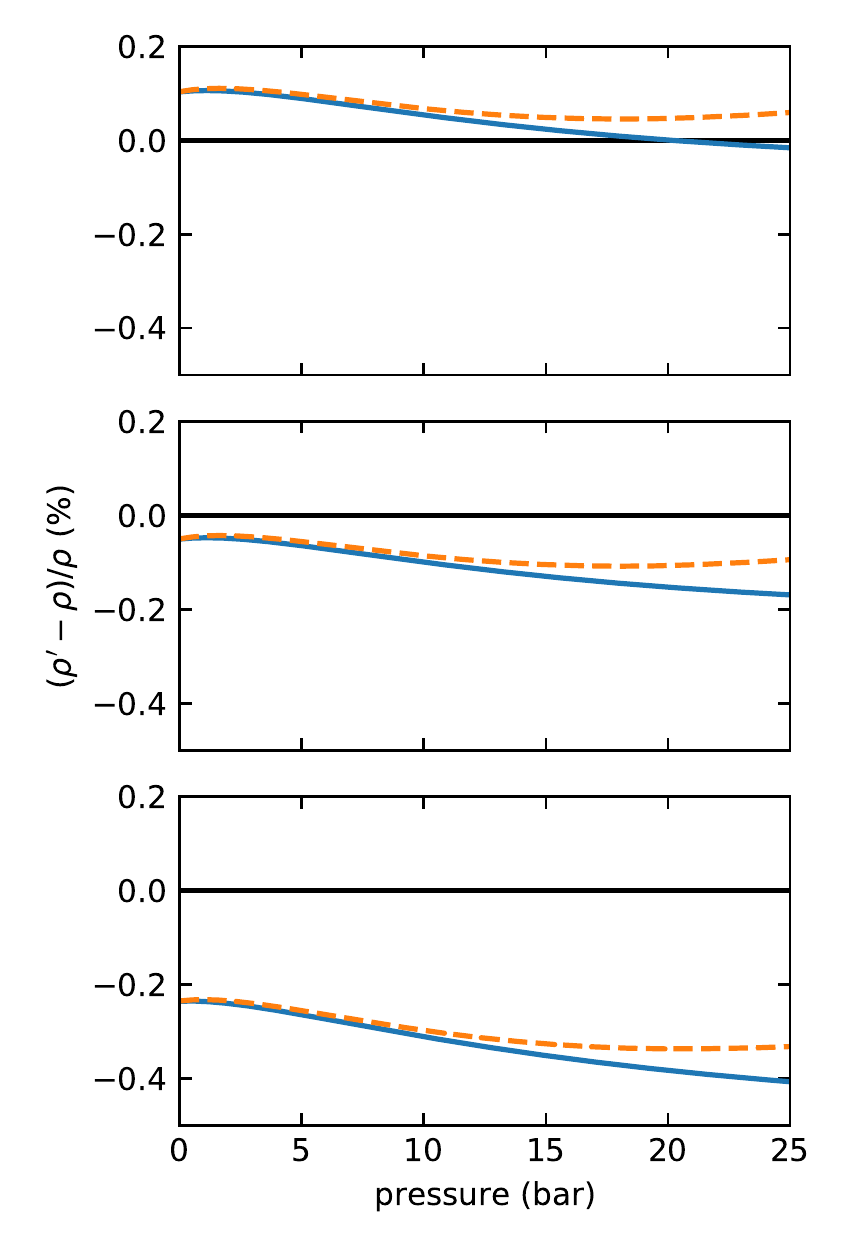}
    \caption{The difference in measured pressure dependence of the density $(\rho' - \rho)/\rho$ where $\rho'$ is either the data of Abraham \emph{et al.} \cite{Abraham1970} (solid blue line) or Tanaka \emph{et al.} \cite{Tanaka2000} (dashed orange line) and $\rho$ is the density obtained here. The three panels show the effect of polarizability on $\rho$ calculated from the dielectric constant \eqref{eq:eps-fit} using the Clausius-Mossotti relation \eqref{eq:clausius-mossotti}. Top panel, $\alpha = 0.1234853$ cm$^{3}$ \cite{Gaiser2018}; middle $\alpha = 0.123296$ cm$^{3}$ \cite{Harris-Lowe1970}; and lower $\alpha = 0.123413 - 0.002376\rho$ cm$^{3}$ (where $\rho$ is in g$\cdot$cm$^{-3}$) \cite{Kerr1970}. Using the polarizability of Ref.~\cite{HEPAK} results in low-pressure deviation reaching almost 1\%.}
    \label{fig:deviations}
\end{figure}

The calculated density crucially depends on the chosen value of polarizability. The pressure dependence of density is tabulated by Brooks and Donnelly \cite{Brooks1977}, who derived values from the functional form reported by Abraham \emph{et al.}~\cite{Abraham1970}. Tanaka \emph{et al.}~\cite{Tanaka2000} later made capacitive measurements of the helium density, producing their own functional form. The comparison of these two past experiments with present data is shown in Fig.~\ref{fig:deviations} using three different values of polarizability $\alpha$. We see that depending on the choice of polarizability, the typical deviations are quite significant and in the range 0.1\% -- 0.5\%. Note, however, that the apparent low-pressure agreement between Abraham \emph{et al.}~\cite{Abraham1970} and Tanaka \emph{et al.}~\cite{Tanaka2000} is artificial, since both of these experiments measured only relative change in density with respect to the zero-temperature, zero-pressure limit value $\rho_0$. In both cases, $\rho_0 = 145.13$~kg~m$^{-3}$ was chosen, which was obtained by Kerr and Taylor \cite{Kerr1964}, who measured changes in density with respect to a reference point near 1.2 K for which an error bar was not specified and then extrapolating a fit below approximately 1 K \cite{Kerr1964}.

While the low-pressure value of $\rho_0$ has a fairly weak empirical basis and uncertainties in the polarizability complicate absolute comparisons, it is clear from Fig.~\ref{fig:deviations} that pressure dependence differs among the experiments. In the present case, the largest uncertainties are likely due to cavity deformation which is not captured accurately enough using elastic compressibility. Another issue might arise in the neglected viscous flow through the heat exchanger in the calculation of the fountain pressure correction \eqref{eq:dp-fountain}. These issues can be mitigated in future experiments, for example, using cryogenic valves and mechanically stronger cavity materials.

\begin{figure}
    \centering
    \includegraphics{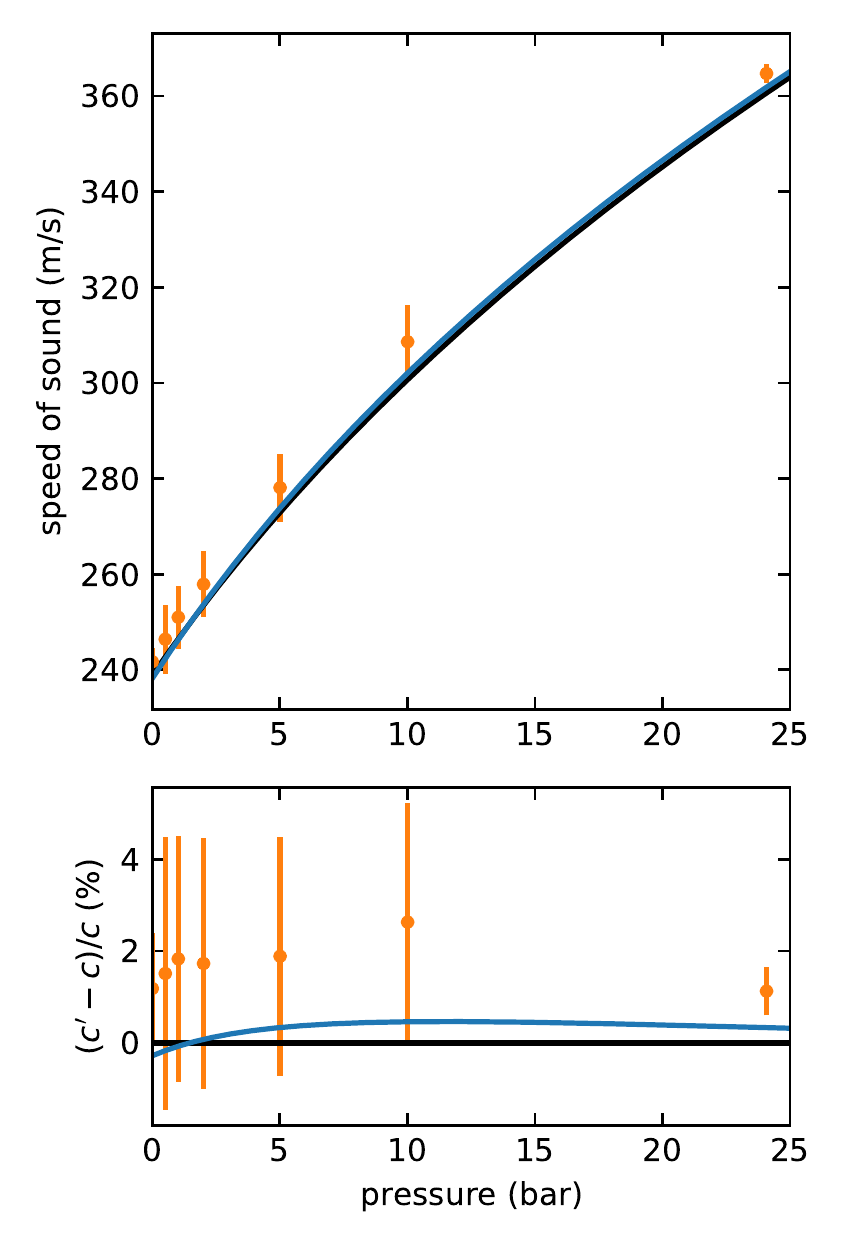}
    \caption{(top) Speed of sound calculated from our pressure dependence of the dielectric constant using \eqref{eq:eps-fit} and polarizability of Ref.~\cite{Gaiser2018} (black line), speed of sound from ultrasound pulse propagation \cite{Abraham1970} (blue line), and from inelastic neutron scattering \cite{Godfrin2021} (orange points). (bottom) Relative deviation of the other datasets ($c'$) from ours ($c$).}
    \label{fig:sound-speed}
\end{figure}

Finally, in Fig.~\ref{fig:sound-speed}(a) we show the speed of sound according to \eqref{eq:sound-speed}. Using the polynomial expression \eqref{eq:eps-fit} and the Clausius-Mossotti equation \eqref{eq:clausius-mossotti} yields
\begin{equation}
    \label{eq:sound-interpolation}
    c = \sqrt{\frac{4\pi\alpha}{3M_4}\left(A_1 + 2A_2\mu + 3A_3\mu^2\right)},
\end{equation}
where to obtain $c(P)$ the expression \eqref{eq:eps-fit} is first inverted to obtain $\mu(P)$. In Fig.~\ref{fig:sound-speed}(b) we show the relative difference between the data obtained here and the speed of sound obtained using ultrasonic pulses by Abraham \emph{et al.}~\cite{Abraham1970} and inelastic neutron scattering by Godfrin \emph{et al.}~\cite{Godfrin2021}. Apart from the low pressure region, our data lie systematically bellow the ultrasound velocities. Since the data reported in Ref.~\cite{Abraham1970} have relative uncertainty of approximately 0.1\% (due to uncertainty of zero-pressure limit value $c(0)$ and statistical uncertainty of the fit), this, again, most likely indicates that the deformation of the cavity is not fully accounted for by linear elastic compressibility.

\section{Conclusions}
We have presented measurements of the dielectric constant and density of superfluid \4He in the zero-temperature limit for pressures up to 25 bar, showing that very high experimental accuracy is attainable under cryogenic conditions using a superconducting microwave cavity. Reviewing multiple past experiments, we find a systematic discrepancy between low-frequency and high-frequency measurements of the dielectric constant -- measured in the present experiment with the highest precision to-date -- which exceeds the expected frequency dependence of the polarizability \cite{Piszczatowski2015}. For the pressure-dependent density, after careful consideration of dominant sources of systematic errors -- a hydrostatic pressure head, fountain effect in the helium fill line, and the cavity compressibility -- we find moderate discrepancies with respect to the values of density reported in the literature, which could be to large extent attributed to the nonlinear deformation of the cavity geometry and rather uncertain value of the molecular polarizability of liquid \4He. Finally, using the measured pressure dependence of density we calculate the speed of sound which is found to be in good agreement, but systematically underestimating, the speed of sound obtained either by ultrasonic pulse propagation \cite{Abraham1970} or inelastic neutron scattering \cite{Godfrin2021}.

The uncertainty of polarizability in the high-density liquid is in stark contrast with the \4He gas near the triple point of water, where experimental accuracy \cite{Gaiser2018} and \emph{ab-initio} calculation of \4He polarizability \cite{Piszczatowski2015} advanced to the point where helium can be used for metrological purposes, such as creation of a pressure standard. Such detailed, quantitative understanding of liquid helium under cryogenic conditions is equally desirable and would allow, for example, accurate calibration of cryogenic secondary pressure transducers. Thanks to high achievable quality factors of superconducting microwave cavities and the high purity of cryogenic liquid helium, extremely accurate measurements of dielectric properties of \4He are possible, which presents an ideal test bed for future extensions of \emph{ab-initio} calculations. Finally, we note that if filled with \3He, a similar system could be used as a highly accurate and sensitive primary thermometer at very low temperatures \cite{Greywall1982}.

\begin{acknowledgments}
This work was supported by the University of Alberta; the Natural Sciences and Engineering Research Council, Canada (Grants RGPIN-2016-04523, CREATE-2017-495446, and  RGPIN-2022-03078); and the Alberta Quantum Major Innovation Fund.  As researchers at the University of Alberta, we acknowledge that we are located on Treaty 6 territory, and that we respect the histories, languages, and cultures of First Nations, M\'etis, Inuit, and all First Peoples of Canada, whose presence continues to enrich our vibrant community.
\end{acknowledgments}


%

\end{document}